\documentclass[preprint2]{aastex6}

\usepackage{color}
\usepackage{mathpazo}
\usepackage{float}
\usepackage{graphicx}
\graphicspath{ {images/} }

\begin{document}

\title{NEW INSIGHTS IN THE SPECTRAL VARIABILITY AND PHYSICAL CONDITIONS OF THE X-RAY ABSORBERS IN NGC 4151}
\author{J. D. Couto\altaffilmark{1}, S. B. Kraemer\altaffilmark{1}, T. J. Turner\altaffilmark{2} 
and D. M. Crenshaw\altaffilmark{3}}

\altaffiltext{1}{Department of Physics, Catholic University of America, Washington, DC 20064}
\altaffiltext{2}{University of Maryland Baltimore County, Baltimore, MD, 21250}
\altaffiltext{3}{Georgia State University, Atlanta, GA, 30302}

\begin{abstract}

We investigate the relationship between the long term X-ray spectral variability in the Seyfert 1.5 galaxy NGC 4151 and its intrinsic absorption, by comparing the 2014 simultaneous ultraviolet/X-Ray observations taken with {\it Hubble STIS Echelle} and {\it Chandra HETGS} with archival observations from {\it Chandra}, {\it XMM-Newton} and {\it Suzaku}. The observations are divided into "high" and "low" states, with the low states showing strong and unabsorbed extended emission at energies below 2 keV.  Our X-ray model consists of a broken powerlaw, neutral reflection and the two dominant absorption components identified by \citet{2005ApJ...633..693K}, hereafter KRA2005, X-High and D+Ea, which are present in all epochs. The model fittings suggest that the absorbers are very stable, with the principal changes in the intrinsic absorption resulting from variations in the ionization state of the gas as the ionizing continuum varies. However, the low states show evidence of larger column densities in one or both of the absorbers. Among plausible explanations for the column increase, we discuss the possibility of an expanding/contracting X-ray corona. As suggested by KRA2005, there seem to be contributions from magnetohydrodynamic (MHD) winds to the mass outflow. Along with the ultra fast outflow absorber identified by \citet{2010A&A...521A..57T}, X-High is consistent with being magnetically driven. On the other hand, it is unlikely that D+Ea is part of the MHD flow, and it is possible that it is radiatively accelerated. These results suggest that at a sufficiently large radial distance there is a break point between MHD-dominated and radiatively driven outflows.

\end{abstract}


\section{Introduction}
\label{intro} 

It is well established that Active Galactic Nuclei (AGN) exhibit extreme X-ray flux variability as they accrete matter onto a supermassive black hole (SMBH). Such variability has been observed in timescales of minutes to hours, up to longer timescales \citep{1999ApJ...524..667T, 2001ASPC..224..167N, 2009A&ARv..17...47T}. Seyfert galaxies correspond to a subclass of AGNs, characterized in terms of  total luminosity coming from their nuclei. The spectrum exhibit broad emission lines that cover a wide range in ionization \citep{1985ApJ...297..166O}, and depending on the characteristics and kinematics presented of such lines, are further subclassified into Type 1 and Type 2, among more specific subclassifications defined by \citet{1976MNRAS.176P..61O} and \citealt{1985ApJ...297..166O}. 

Furthermore, mass outflows, evidenced by blueshifted absorption lines of ionized gas coming from the central regions of Seyfert galaxies \citep{2003ARA&A..41..117C}, are believed to be closely related to AGN spectral power. They play an important role in understanding the structure, dynamics and evolution of AGNs, and therefore are fundamental in order to investigate the extent of the contribution of outflows to AGN Feedback. Although the origin of X-ray variability is still not fully understood, in a general way the spectral changes can be attributed either to time dependent structural changes in the absorption features or source variations, and sometimes it is very difficult to narrow down a single solution to model the observed spectrum. Many different scenarios have been proposed to account for spectral variability in AGNs, and below we list a few of them:

\renewcommand{\labelenumi}{\roman{enumi}}
\begin{enumerate}

\item{The X-ray source varies in luminosity, due to changes in the mass accretion rate, structural changes in the accretion disk, flaring events, etc, causing the intrinsic continuum flux to vary accordingly, and the absorption components to respond by changing their ionization state, hence their opacity to the incident radiation; this scenario implies a certain stability in the nature of the X-ray absorbers, which has been previously discussed by \citet{2014ApJ...797..105S} in the study of NGC 3783.} 

\item{The X-ray source remains roughly constant in luminosity or varies on a much longer timescale, and the observed variability appears when high column density clouds of intervening gas temporarily cover the source and shield the absorbers, causing their ionization state to vary, as suggested in \citet{2013ApJ...762...80T}. Specific examples include MGC $-$6$-$30$-$15 (\citealt{1998ApJ...501L..29M}), and also NGC 3516 (\citealt{2011ApJ...733...48T}; in which their Figure 1 illustrates the extreme variability in the absorption profile).}

\item{Major changes in the spectrum are not due to variations in the X-ray source, which may vary or remain constant, but to changes in column density of the absorbers due to transverse motion of gas, as proposed i.e. by KRA2005 for the 2002 and 2000 {\it Chandra} spectra on NGC 4151, and more recently by \citet{2014Sci...345...64K} for NGC 5548. Spectral variability could also be attributed to changes in the relative fractions of intrinsic and reflected continuum. as \citet{2010MNRAS.403..196M} discuss in the study of NGC 4051, such changes can be explained by effects of reverberation lags, caused by high covering factor reflecting material.}

\end{enumerate}

All of the above scenarios make physical sense and have been probed with observational evidence in studies of different AGNs. The idea of single scenario to account for the X-ray variability still does not seem plausible, and conclusions must rely on the data available for each individual galaxy. 

In this paper we will focus on investigating the long term flux variability and intrinsic absorption of NGC 4151 on a timescale of 14 years. \objectname[NGC]{NGC 4151} is one of the nearest (z=0.0033; \citealt{1991rc3..book.....D}) and brightest AGN and is classified as a Seyfert 1.5, exhibiting comparable broad and narrow emission components in H$\beta$ \citep{1985ApJ...297..166O}. Over the past decades, NGC 4151 has been the target of many single and multiwavelength observations, and its heavily absorbed X-ray spectrum and complex intrinsic absorption features have been extensively studied. As discussed by \citet{2007ApJ...659..250C}, the complexity of the intrinsic absorption in NGC 4151 could be related with the 45$^{\circ}$ viewing angle \citep{2005AJ....130..945D} of the black hole/accretion disk system. The hard X-ray (2$-$10 keV) shows strong flux variability and the soft energy band ($<$ 2 keV) is dominated by emission lines with velocity widths that would indicate that they are coming from the intermediate line region (ILR), located at $\sim$ 1 pc, up to the narrow line region (NLR), that can extend from $\sim$~10~pc~to $\sim$~1 kpc \citep{2011ApJ...729...75W}.

As we will discuss in more detail in this paper, our initial assumption is that the X-ray flux variability is related to the opacity of the gas, although KRA2005 found no simple correlation between the ionization state and the intrinsic continuum flux, and pointed out that this result corroborates with the idea of transverse motion of gas across our line-of-sight. Furthermore, our analysis will provide us with insights in the temporal evolution of the X-ray absorbers and the nature of the mass outflows in NGC 4151. Section 2, we give a detailed description of the observations and data reduction; Section 3, we discuss about the X-ray spectral variability and the X-ray model; Section 4, we present our results and comparison between different epochs; in Section 5 we will discuss the implications of our analysis, and finally Section 6 gives a summary and final conclusions.

\floattable
\begin{deluxetable}{lcccccc}
\tablenum{1}
\tabletypesize{\small}
\tablecaption{Observation Log}
\tablewidth{0pt}
\tablehead{
\colhead{ } & \colhead{Obs ID} & \colhead{Start Time} & \colhead{Instrument} &
\colhead{Duration} & \colhead{Flux$_{2-10 keV}$} & \colhead{Flux State} \\
\colhead{ } & \colhead{ } & \colhead{(UTC)} & \colhead{ } &
\colhead{(ks)} & \colhead{(10$^{-11}$ erg cm$^{2}$ s$^{-1}$)} & \colhead{ }
}
\startdata
\it{Chandra}    & 335       & 2000-03-05 23:19:11 & HETG/ACIS-S & 47.4  & $6.41 \pm 0.09$ & Low  \\
                & 3052      & 2002-05-09 18:26:49 & HETG/ACIS-S & 153.1 & $16.7 \pm 0.02$ & High \\
                & 3480      & 2002-05-07 23:25:58 & HETG/ACIS-S & 90.8  &                 & High \\  
                & 3089      & 2002-07-02 15:56:56 & LETG/ACIS-S & 86.3  & $15.2 \pm 0.02$ & High \\
                & 7830      & 2007-07-21 10:10:28 & HETG/ACIS-S & 49.3  & $10.6 \pm 0.02$ & High \\
                & 16089     & 2014-02-12 19:47:34 & HETG/ACIS-S & 171.9 & $6.27 \pm 0.03$ & Low  \\
                & 16090     & 2014-03-08 15:20:33 & HETG/ACIS-S & 68.8  &                 & Low  \\
\hline
\it{XMM-Newton} & 112830201 & 2000-12-22 10:42:18 & EPIC-PN     & 47.1  & $4.69 \pm 0.01$ & Low  \\
                & 112830501 & 2000-12-22 02:53:42 & EPIC-PN     & 11.2  &                 & Low  \\
\hline
\it{Suzaku}     & 701034010 & 2006-12-18 20:05:09 & XIS,HXD     & 155.8 & $4.30 \pm 0.01$ & Low  \\
\enddata
\end{deluxetable}

\section{Observations and Data Reduction}
\label{obs}

Our analysis will be focused on the 2014 observation of \objectname[NGC]{NGC 4151}, part of a simultaneous ultraviolet (UV) and X-ray program using, respectively, {\it Hubble Space Telescope}'s Space Telescope Imaging Spectrograph (STIS) Echelle and {\it Chandra X-ray Observatory} High-Energy Transmission Grating Spectrometer (HETGS). The {\it Chandra} observation sums a total of ~241 ks in exposure time, and was performed in two epochs due to constraints in the roll angle alignment to match the cross-dispersion direction of the extended emission of NGC 4151 \citep{2000ApJ...545L..81O}. The first data (ObsID 16089, ~172 ks) was obtained over the period 2014 February 12 19:47 to 2014 February 14 21:04 UTC, and the second (ObsID 16090, ~69 ks) on 2014 March 08 15:20 to 2014 March 09 11:26 UTC. Both datasets were obtained with the \arcsec S-array\arcsec of the Advanced CCD Imaging Spectrometer (ACIS$-$S). CCDs S1 to S5 were read out.

We also include in our analysis a set of multisatellite archival observations of NGC 4151 from {\it Chandra}, {\it XMM-Newton} and {\it Suzaku}. The details for each individual observation can be found in Table 1. We classified the epochs presented in this paper in terms of their flux state as high (for fluxes $>$ 1.0 $\times$ $10^{-10}$ ergs cm$^2$ s$^{-1}$) and low (for fluxes $<$ 1.0 $\times$ $10^{-10}$ ergs cm$^2$ s$^{-1}$).

\subsection{Data Reduction}
All {\it Chandra} datasets were processed following the {\it Chandra} ACIS science threads, with CIAO (version 4.6) software, and HEASOFT (version 6.15.1) packages. The calibration files from {\it Chandra} CALDB (version 4.5.9) were used. The HETGS is composed of two grating arms $-$ the High Energy Grating (HEG), and the Medium Energy Grating (MEG). HEG and MEG are placed in such a way that the dispersed spectra from the two are perpendicular on the chip, appearing as an X around the target, and combined they cover from 0.4$-$10 keV in energy range. The Low-Energy Transmission Grating Spectrometer (LETGS) is composed of a single arm, the Low Energy Grating (LEG), and it is optimized for lower energies ($<$~2~keV). The script {\it chandra\_repro} automatically runs all the standard data processing steps, such as destreak at the ACIS-CCD8 chip, reprocessing of bad pixels file and level=2 event and pulse height amplitude files (PHA), filtering of good time intervals, and creation of response matrix files (RMF) and ancillary response files (ARF) for positive and negative grating arms. The background was negligible, so there was no need for background extraction. The $\pm $~first-orders PHA, RMF and ARF of each dataset were combined, and all spectra were rebinned to a minimum of 30 counts per bin, except for the {\it{Chandra}} 2000 (Obs ID 335) and 2007 (Obs ID 7830) datasets, which were rebinned to a minimum of 15 counts per bin due to shorter exposure time. The total number of counts for the 2014 HEG spectra was 26931 for Obs ID 16089 and 9246 for Obs ID 16090. For the 2014 MEG spectra was 24025 for Obs ID 16089 and 8115 for Obs ID 16090. For convenience, some of the observations were either combined ({\it Chandra} 2002 $-$ ObsID 3480 and 3520, and 2014 $-$ ObsID 16089 and 16090) or fitted together ({\it XMM-Newton} 2000 $-$ ObsID ...~201 and ...~501) in order to increase the number of counts per bin, and improve the fitting statistics. The total number of counts in the combined spectrum was 35938 counts, with a count rate of 0.1483 $\pm$ 0.0010 counts $s^{-1}$ for HEG, and 32115 counts, with a count rate of 0.1326 $\pm$ 0.0009 counts s$^{-1}$ for MEG.

The XMM-Newton datasets were processed with the XMM-Newton Science Analysis System (SAS) (version 14.0.0) and calibration files as of 2015 February. The EPIC-PN spectra were time and rate filtered, and limited to single and double events (PATTERN=0$-$4). Events near bad pixels and edges were also excluded (FLAG==0). The source and background regions were defined in detector coordinates as a circle with radius of 640 and 740 detector pixels, respectively. The pile-up impact was negligible on the selected datasets. The redistribution matrix and ancillary response files were generated with the XMM-SAS tasks {\it rmfgen} and {\it arfgen}, so the data set would be ready to be fitted with XSPEC.

The {\it Suzaku} observation  was taken with both X-ray Imaging Spectrometer (XIS) and the Hard X-ray Detector (HXD). Among the selected observations, this spectrum covers the largest observable energy range, from 0.6$-$50 keV. The data reduction was performed as described in the Observation and Data Reduction section in \citet{2016arXiv160707125Y}.

\begin{figure*}
\figurenum{1}
\epsscale{1.05}
\plotone{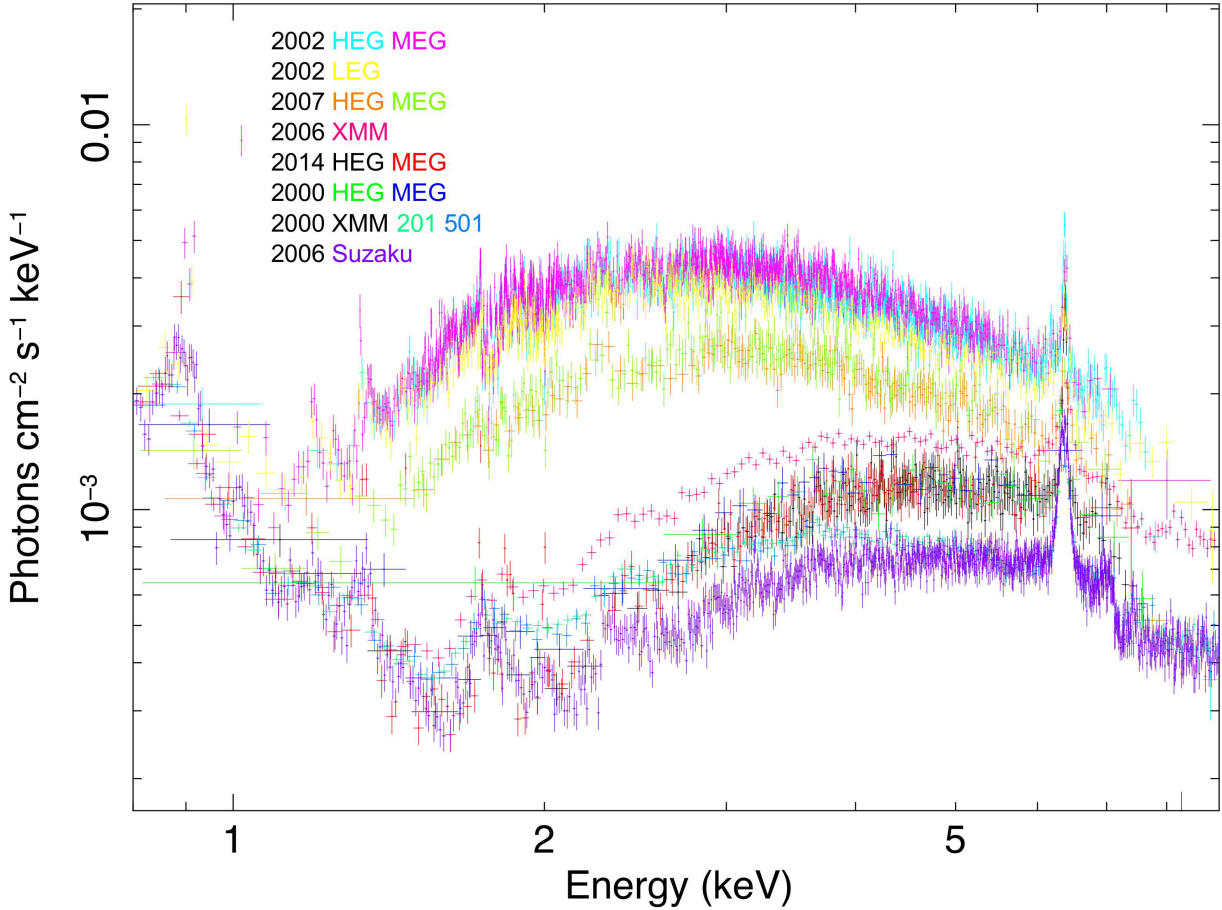}
\caption{X-ray Spectra of NGC 4151 from 2000 to 2014. The observations are as follows, 2014 {\it Chandra} HEG (black) and MEG (red); 2000 {\it Chandra} HEG (dark green) and MEG (dark blue); 2002 {\it Chandra} HEG (cyan), MEG (magenta) and LEG (yellow); 2007 {\it Chandra} HEG (orange) and MEG (light green); 2000 {\it XMM-Newton} ObsID ...201 (green+cyan) and ...501 (blue+cyan); and 2006 {\it Suzaku} (purple). For illustration purposes we show in pink the 2006 {\it XMM-Newton} ObsID 0402660201, since it will be discussed in further sections. The datasets were extracted with 15 or 30 counts per bin, depending on the total number of counts of the observation.\label{fig:fig1}}
\end{figure*}

\begin{figure}
\figurenum{2}
\epsscale{1.1}
\plotone{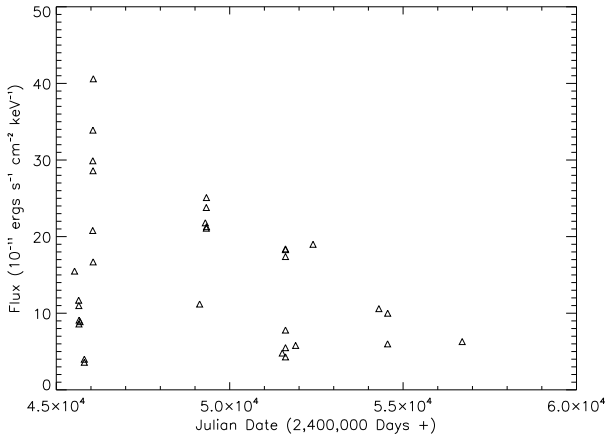}
\caption{X-ray light curve continuum of NGC 4151 from 1983 to 2014. 2-10 keV fluxes are plotted as a function of Julian date.
\label{fig:fig2}}
\end{figure}

\begin{figure}
\figurenum{3}
\epsscale{1.1}
\plotone{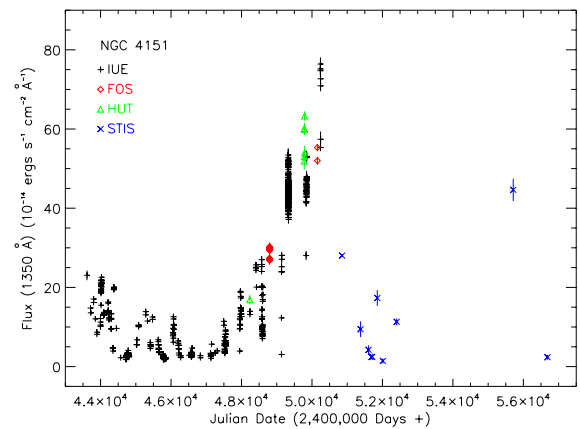}
\caption{UV light curve continuum of NGC 4151 from 1976 to 2014. Fluxes at 1350 {{\AA}} are plotted as a function of Julian date. The symbols correspond to different instruments: pluses, IUE; triangles, FOS; diamonds, HUT; crosses, STIS. Vertical lines indicate the error bars for 1$\sigma$.
\label{fig:fig3}}
\end{figure}

\section{X-ray Spectral Variability and Modeling}
\label{allspec}

All of the X-ray data analysis and model fitting were performed with the High Energy Astrophysics Science Archive Research Center (HEASARC) tools package HEASOFT (version 6.15.1), and XSPEC (version 12.8.1g), \citep{1996ASPC..101...17A}, using $\chi^2$ statistics. Model parameters are given at redshift z~=~0.0033, and errors are calculated on XSPEC for 90$\%$ confidence limits. Galactic absorption by dust and gas from our galaxy is represented by a full-coverage column of gas with fixed hydrogen column density, N$_{H}$~=~1.98~$\times$~$10^{20}$ $cm^{-2}$, defined from the dust map in \citet{1990ARA&A..28..215D}.

The X-ray spectral variability of \objectname[NGC]{NGC 4151} over the past 14 years is illustrated in Figure 1, which shows the unfolded spectra as a function of energy for the datasets described in the previous section. The lower flux state datasets show a more heavily absorbed spectrum than the higher flux state ones, however, it is also worth noting the similarities in spectral shape and continuum flux between different epochs. More specifically, it is remarkable how the low flux states 2000 and 2014 {\it Chandra} HETG, and high flux states 2002 HETG and LETG {\it{Chandra}} observations overlap. Furthermore, the X-ray lightcurve over the past $\sim$~30 years is illustrated in Figure 2. The 2$-$10 keV flux for the 2014 {\it{Chandra}} spectra was $\sim$ 6.27 $\times$ $10^{-11}$ ergs cm$^2$ s$^{-1}$, roughly a factor of $\sim$~3 times fainter than the 2002 {\it{Chandra}} observation. Interestingly, the flux of the Fe k$\alpha$ emission line at 6.4 keV only varied by a factor of $\sim$~2, which could indicate that the reflection component may remain somewhat constant throughout the years, and is consistent with it being outside the BLR, since the broad lines vary strongly with the continuum \citep{1996ApJ...470..322C}. 

We also observe a factor of $\sim$ 4 decrease in the UV spectrum, with a flux at 1350 {{\AA}} of $\sim$~2.5~$\times$~$10^{-14}$ ergs~cm$^2$~s$^{-1}$~{{\AA}}$^{-1}$. Figure 3, previously published on \citet{2006ApJS..167..161K}, hereafter KRA2006, shows a UV light curve at 1350~{{\AA}} for NGC~4151 over the past $\sim$~30 years, with the last point corresponding to our new 2014 {\it STIS} observation. The notable UV short and long term flux variability rules out the possibility of existence of clouds of intervening gas in NGC~4151, as described in scenario (ii) of the Introduction. For the UV to vary due temporary occultation, the intervening gas must be dusty, which is very unlikely given the fact that the gas should be located in the inner part of the BLR, hence within the dust sublimation radius ($\sim$~0.05 pc, from \citealt{1987ApJ...320..537B} Eq. 5 for T~=~1500 K and L$_{\nu}$~$\sim$~10$^{43}$ erg s$^{-1}$; see KRA2006). Thus, spectral changes would be either caused by transverse motion or by changes in the ionization state of the absorbing gas.

\begin{figure}
\figurenum{4}
\epsscale{1.2}
\plotone{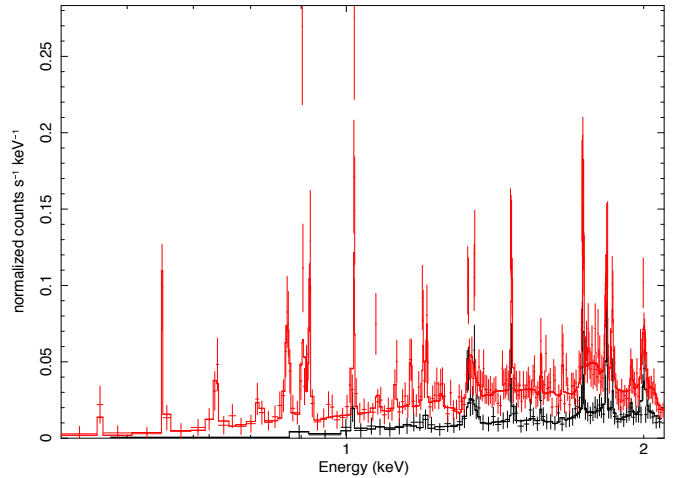}
\caption{Soft X-Ray Emission for the combined HEG (black) and MEG (red) 2014 Chandra spectrum for the lines listed in Table 2.
\label{fig:fig4}}
\end{figure}

\begin{figure}
\figurenum{5}
\epsscale{1.2}
\plotone{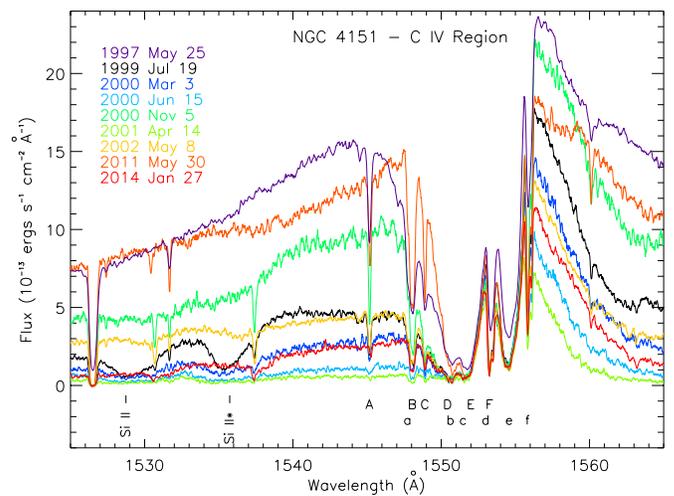}
\caption{STIS spectra of NGC 4151 in the \ion{C}{4} region obtained between 1997 and 2014. The locations of the kinematic components defined by \citet{1997ApJ...483..717W} are labeled in capital letters for \ion{C}{4} $\lambda$1548 and lowercase letters for \ion{C}{4} $\lambda$1551 {\it (see \citealt{2006ApJS..167..161K})}.
\label{fig:fig5}}
\end{figure}

\subsection{X-ray Modeling}

Our approach was to model the 2014 dataset based on the model parameters proposed by KRA2005. The main features, such as absorption and reflection components, and the soft X-ray region, will be discussed in more detail in the following subsections. The hard X-ray model can be summarized as follows:

\begin{itemize}
\item []{\it{Broken powerlaw}}. The powerlaw is parameterized in terms of the photon spectral indices ($\Gamma$$_{1}$ and $\Gamma$$_{2}$) and normalization parameter, with an energy break at 0.5 keV, representing the intrinsic X-ray continuum emission.  $\Gamma$$_{1}$ is fixed at 2.3 for energies below 0.5 keV, and $\Gamma$$_{2}$ is fixed at 1.5. Although $\Gamma_{2}$ is flatter than typical (see \citealt{1998ApJS..114...73G}, and references therein), it is consistent with previous studies of NGC 4151. Investigating the cause of the flat photon index would be out of the scope of this paper, given the limitations in energy range in our datasets. 

\item []{\it{Reflection component + Fe K$\alpha$ emission line}}. Neutral Compton reflection that arises from the reprocessing of X-rays from cold matter. The reflected component has the same spectral shape as the incident continuum in terms of photon index and normalization parameters, and it is described in more detail in the following subsections. 

\item []{\it{Dual absorption component}}. The absorbers will be modeled in terms of changes in column density and/or ionization parameter, and as in the KRA2005 model we assume that they are radially distributed around the central source. We can identify two main absorption components, with the nearest, more highly ionized one (X-High), filtering the ionizing radiation incident on the less ionized component (D+Ea). \footnote{D+Ea $-$ defined by \citet{2001ApJ...551..671K}. This component is a combination of components D and E, identified by \citet{1997ApJ...483..717W} nomenclature for the Ultraviolet (UV) kinematic components; it can be detected both in the UV and X-ray spectrum.} The absorbers lie in front of the source intrinsic and reflected continuum, and are seen as imprinted absorption features. We start with the assumption that the column densities of the absorbers are equal, and see if we can find a good statistical solution by varying ionization parameter, or if we need to adjust them.

\item []{\it{Emission Lines}}. The spectrum is dominated by emission lines in the soft X-ray band, below 2~keV, as illustrated in Figure 4. The emission lines are represented in the model as gaussians, and correspond to the extended emission of NGC 4151. The lines were based on the soft X-ray emission of \objectname[NGC]{NGC 1068} {\it{Chandra}} spectrum presented in \citet{2014ApJ...780..121K}, and are summarized on Table 2 (note that {\it Chandra} HETG has a spectral resolution of $\Delta$E~=~0.4$-$77 eV FWHM and is capable of resolving FWHM down to $\sim$ 300 km s$^{-1}$). The physical conditions and kinematics of the extended emission region will be studied in more detail in a future paper (Couto et. al, in preparation).
\end{itemize}

\floattable
\begin{deluxetable}{cccccc}
\tablenum{2}
\tabletypesize{\small}
\tablecaption{NGC 4151 Emission Lines}
\tablewidth{0pt}
\tablehead{
\colhead{Wavelength} & \colhead{Energy} & \colhead{Flux} & \colhead{Lab} & \colhead{Ion} & \colhead{FWHM} \\
\colhead{($\AA$)} & \colhead{(keV)} & \colhead{($cm^{-2}$$s^{-1}$)} & \colhead{($\AA$)} & \colhead{ } & \colhead{(km $s^{-1}$)}
}
\startdata
6.178  & 2.007 & $7.24^{+2.72}_{-2.22}$$\times$$10^{-6}$ & 6.180  & \ion{Si}{14} & $\geq$ 1400 \\
6.647  & 1.866 & $7.68^{+1.78}_{-1.78}$$\times$$10^{-6}$ & 6.640  & \ion{Si}{13} & $\geq$ 1600 \\
6.692  & 1.853 & $9.02^{+0.37}_{-0.37}$$\times$$10^{-7}$ & 6.690  & \ion{Si}{13} & $\sim$ 1600 \\
6.739  & 1.840 & $8.43^{+1.53}_{-1.53}$$\times$$10^{-6}$ & 6.744  & \ion{Si}{13} & $\sim$ 1700 \\
6.988  & 1.774 & $7.24^{+2.72}_{-2.22}$$\times$$10^{-6}$ & 7.001  & \ion{Si}{8} (K$\alpha$) & $\geq$ 1400 \\
7.113  & 1.743 & $6.50^{+0.82}_{-1.35}$$\times$$10^{-6}$ & 7.110  & \ion{Mg}{12} & $\geq$ 800 \\
7.466  & 1.661 & $6.50^{+0.82}_{-1.35}$$\times$$10^{-6}$ & 7.110  & \ion{Mg}{12} & $\geq$ 800 \\
7.848  & 1.580 & $1.43^{+1.03}_{-1.03}$$\times$$10^{-6}$ & 7.850  & \ion{Mg}{11} & $\sim$ 450 \\
8.404  & 1.476 & $5.71^{+1.49}_{-1.49}$$\times$$10^{-6}$ & 8.420  & \ion{Mg}{12} & $\geq$ 1000 \\
9.070  & 1.367 & $1.29^{+2.42}_{-1.29}$$\times$$10^{-6}$ & 9.170  & \ion{Mg}{11} & $\geq$ 2600 \\
9.244  & 1.341 & $2.17^{+4.74}_{-4.74}$$\times$$10^{-6}$ & 9.230  & \ion{Mg}{11} & $\geq$ 4000 \\
9.898  & 1.253 & $1.21^{+2.11}_{-1.21}$$\times$$10^{-6}$ & 9.800  & \ion{Fe}{17} (rrc) & $\sim$ 700 \\
10.248 & 1.210 & $4.52^{+2.29}_{-2.29}$$\times$$10^{-6}$ & 10.240 & \ion{Ne}{10} & $\leq$ 700 \\
10.323 & 1.201 & $6.27^{+3.01}_{-3.01}$$\times$$10^{-6}$ & 10.388 & \ion{Ne}{9}  & $\sim$ 1600 \\
10.635 & 1.166 & $4.61^{+2.66}_{-2.66}$$\times$$10^{-6}$ & 10.641 & \ion{Fe}{19} & $\leq$ 800 \\
11.031 & 1.124 & $3.35^{+3.25}_{-5.15}$$\times$$10^{-6}$ & 11.000 & \ion{Ne}{9}  & $\sim$ 1600 \\
11.495 & 1.079 & $4.71^{+5.15}_{-4.71}$$\times$$10^{-6}$ & 11.500 & \ion{Fe}{18} & $\sim$ 3300 \\
12.144 & 1.021 & $2.50^{+0.62}_{-0.62}$$\times$$10^{-5}$ & 12.100 & \ion{Ne}{10} & $\geq$ 1100 \\
13.457 & 0.921 & $3.53^{+1.06}_{-1.06}$$\times$$10^{-5}$ & 13.447 & \ion{Ne}{9}  & $\sim$ 1600 \\
13.685 & 0.906 & $5.57^{+1.66}_{-1.66}$$\times$$10^{-5}$ & 13.700 & \ion{Ne}{9}  & $\geq$ 4000 \\
14.169 & 0.875 & $6.40^{+1.74}_{-1.74}$$\times$$10^{-5}$ & 14.250 & \ion{O}{7}   & $\leq$ 4000 \\
15.159 & 0.818 & $2.54^{+2.11}_{-2.11}$$\times$$10^{-5}$ & 15.188 & \ion{O}{8}   & $\sim$ 4500 \\
16.765 & 0.740 & $8.12^{+4.11}_{-4.11}$$\times$$10^{-5}$ & 16.777 & \ion{O}{7}   & $\sim$ 4800 \\
18.892 & 0.653 & $1.24^{+0.57}_{-0.57}$$\times$$10^{-4}$ & 18.968 & \ion{O}{8}   & $\sim$ 1100 \\
21.963 & 0.565 & $4.63^{+2.58}_{-2.58}$$\times$$10^{-4}$ & 21.804 & \ion{O}{7}   & $\geq$ 6500 \\
\enddata
\tablecomments{Large FWHM may be due to blending of lines. Details will be discussed in Couto et. al, in preparation.}
\end{deluxetable}

\subsection{Reflection Component}

We account for Compton reflection on XPEC by using PEXMON model \citep{2007MNRAS.382..194N}, parameterized in terms of the powerlaw intrinsic continuum, with photon index ($\Gamma$) and normalization parameters tied to the primary broken powerlaw, high energy cutoff fixed at 100 keV, solar abundances, relative reflection fraction R, and inclination angle fixed at 45$^{\circ}$. The reflection fraction R is the only parameter allowed to vary, and it represents the ratio between the observed and expected reflection from a slab of gas subtending a solid angle of 2$\pi$R. Negative values of R in the models indicate that we are taking into account only the direct reflected continuum. As previously stated, PEXMON only accounts for reflection from neutral matter. The Fe K$\alpha$ Compton shoulder is also included in PEXMON as a gaussian with energy fixed at 6.315 keV and $\sigma$ equal to 0.035 keV.

The Fe K$\alpha$ equivalent width was $\sim$~170 eV for the 2014 combined {\it{Chandra}} dataset and $\sim$~81 eV for the 2002 combined {\it{Chandra}}; the full width half maximum (FWHM) was $\sim$~2500 km $s^{-1}$ and $\sim$~2080 km $s^{-1}$, for 2014 and 2002, respectively. The Fe K$\alpha$ emission line is clearly detected, specially in the low flux states, and its narrow profile indicates that it is coming from a more distant region, such as the Intermediate Line Region (ILR) \citep{2010ApJS..187..581S}.   

\subsection{Photoionization Models}

The photoionization models for the absorption components were computed using CLOUDY (version 13.00; \citealt{2013RMxAA..49..137F}). The X-ray absorbers are modeled as single-zoned slabs with constant density atomic gas, illuminated by the central source. We assumed roughly solar elemental abundances from \citet{1989AIPC..183....1G} with the following log values, relative to H: He:~$-$~1.00; C:~$-$~3.47; N:~$-$~3.92; O:~$-$~3.17; Ne:~$-$~3.96, Na:~$-$~5.76; Mg:~$-$~4.48; Al:~$-$~5.55; Si:~$-$~4.51; S:~$-$~4.82; Ar:~$-$~5.60; Ca:~$-$~5.66; Fe:~$-$~4.4; Ni:~$-$~5.78. The models are parameterized in terms of the total hydrogen column density N$_{H}$ and ionization state of the gas, given by the dimensionless ionization parameter U~=~Q/(4$\pi$$r^2$c$n_{H}$), in terms of Q, the number of ionizing photons~$s^{-1}$ emitted the source, $r$, radial distance of the absorber, and $n_{H}$ the hydrogen number density. The spectral grids generated with CLOUDY were used as input to XSPEC in the form of {\it{mtables}} (for details in the CLOUDY/XSPEC interface, see \citealt{2006PASP..118..920P}).

For a given range of values of U and N$_{H}$ in log scale, we generated a grid of photoionization models for each of the absorption components. The first, highly ionized absorber (X-High) is assumed to be closest to the source and directly illuminated by the ionized continuum. D+Ea is at larger radial distance, and is illuminated by the radiation filtered by X-High, by selecting the transmitted continuum flux of X-High at 0.952 Ryd as incident continuum of D+Ea, given that at such energy there should not be any significant absorption. As noted above, the column density $N_{H}$ is kept roughly constant for both absorbers, based on the KRA2005 model results. KRA2005 found evidence for five distinct absorption components, namely in order of radial distance: X-High, D+Ea, D+Eb, D+Ec, and E', with the three latter being lower in ionization, and with smaller column density than the first two. However, our models did not show statistical improvement by adding the less ionized components.

Figure 5, (from KRA2006, with the addition of more recent data, including the new 2014 {\it STIS} spectrum), illustrates the C IV intrinsic absorption variability in NGC 4151 from 1997 to 2014; the kinematic absorption component letters are highlighted in the figure for both \ion{C}{4} $\lambda$1548 and $\lambda$1551. Since we see no variations in velocity shift for D+E in the UV spectrum (see Figure 5), the outflowing velocities for the absorption components were assumed to be the same as KRA2005, ~$\sim$~500~km~s$^{-1}$. In our initial modeling we concentrated on fitting the broad-band absorption profile, hence did not consider turbulence. Once an acceptable fit was obtained, we included turbulence to optimize the modeling of individual absorption lines (when detectable).

\section{Results}

Since the atomic data used in CLOUDY have improved considerably in the past decade, we started our analysis by refitting the KRA2005 model to the 2002 dataset. Our model fittings were performed simultaneously on HEG and MEG, constrained to the 2$-$8 keV energy range. The free parameters are column density, ionization parameter and reflection fraction R, while the powerlaw photon index and the gaussian lines for the soft X-ray emission were held fixed. We were able to reproduce the KRA2005 model with almost identical parameters and good statistical significance by fitting the two main X-ray absorption components. The column densities of the absorbers were roughly equal, $N_{H}$~$\sim$~3.2~$\times$~10$^{22}$ cm$^{-2}$, and the ionization parameters were $\log$~U$_{X-High}$~$\sim$~1.09 and $\log$~U$_{D+Ea}$~$\sim$~$-$~0.27. Figure 6 shows the 2002 fit and the model vs. data ratio.

In order to determine whether the individual absorption lines detected by KRA2005 were accurately predicted by our models, we reran X-High and D+Ea photoionization models with the energy resolution increased by a factor of 10 (\citealt{2006PASP..118..920P}). Without turbulence, the predicted lines were significantly shallower than observed. Therefore, we included turbulence of 100 km~s$^{-1}$ (see KRA2005), and the resulting fit in the energy range of the silicon inner shell lines is shown in Figure 7.

\begin{figure}
\figurenum{6}
\plotone{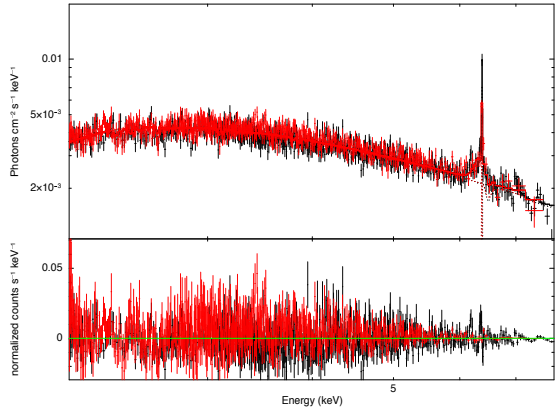}
\caption{Unfolded spectrum and model residuals for the 2002 {\it Chandra} HEG (black) and MEG (red) combined dataset with the KRA2005 model parameters.
\label{fig:fig7}}
\end{figure}

\begin{figure}
\figurenum{7}
\epsscale{1.2}
\plotone{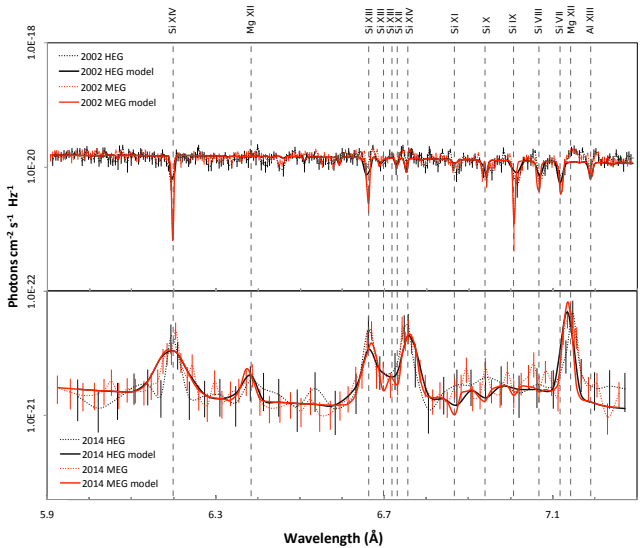}
\caption{Intrinsic absorption features in the 1.7 $-$ 2.1 keV Soft X-ray region of the 2002 and 2014 {\it Chandra} spectra. HEG (black) and MEG (red) data are plotted as dashed lines; solid lines correspond to the model results previously discussed in the text. The strongest absorption and emission features are highlighted in the plot.
\label{fig:fig7}}
\end{figure}

\begin{figure}
\figurenum{8}
\epsscale{1.0}
\plotone{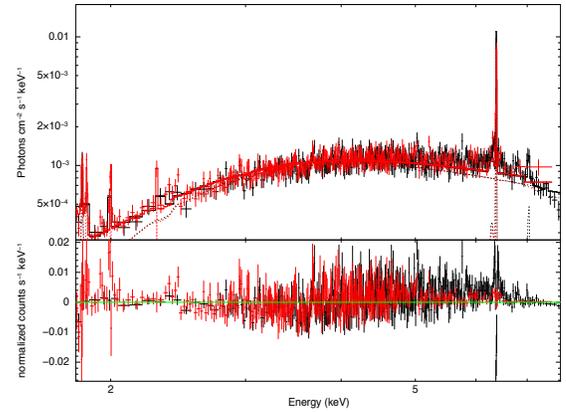}
\caption{HEG (black) and MEG (red) unfolded spectra and model residuals for the best fitting model for the 2014 {\it Chandra} observation. 
\label{fig:fig8}}
\end{figure}

\subsection{Fitting the 2014 {\it Chandra} Observations}

Based on our refitting of the 2002 dataset, we proceeded with the 2014 {\it Chandra} observations. In a lower flux state, all model components play a much important role in shaping the spectrum we observe. The reflected continuum becomes more dominant, and many of features imprinted in the absorption continuum of high flux state epochs are only seen as emission lines in the low flux ones. Without allowing XSPEC to fit the data, we got an acceptable solution, identical to the KRA2005 fitting for the 2000 dataset, which suggested transverse motion manifested by a considerable increase in the column density of D+Ea ($N_{H}$~$\sim$~8~$\times$~10$^{22}$ cm$^{-2}$). However, in allowing the parameters of D+Ea to vary, the fitting favored an unphysical solution in which U$_{D+Ea}$ increased, which seemed highly unlikely given its lower flux state.

Thus, following the fitting of the 2002 dataset we tied column densities of X-High and D+Ea. It was evident in the fittings, though, that a bigger column was needed to properly model the low flux spectrum. We were able to reproduce the 2014 spectrum shape, and a final satisfactory statistical fit was achieved when we allowed a fraction of the intrinsic continuum to be seen without any attenuation, due, for example, to scattering by free electrons. Such component is modeled as an uncovered powerlaw with photon index $\Gamma$~$\sim$~2.4, contributing mostly to the soft energy range of the spectrum. Our best model fitting for the 2014 data is illustrated in Figure 8. Variations in the gas opacity are still a major contributor to the observed spectral changes between high and low states, but we also noticed an increase in the 2014 column density when compared with the 2002 model ($N_{H}$~$\sim$5~$\times$~10$^{22}$ cm$^{-2}$). Such values for column density imply in the saturation of predicted ionic column densities of low ionization ions such as C II, Si II and O~I (see Table 3). The uncovered scattering is evidenced in the lower flux state, where it contributes more significantly to the total flux. However, by refitting the 2002 data to include the scattering component, we obtain an improvement of 0.1 in $\chi$$^{2}$ in the 0.5$-$8.0 keV range. 

Figure 7 also shows a a comparison between the 2002 and 2014 intrinsic absorption in the 1.7~$-$~2.1 keV band. Some of the emission lines coming from the extended emission region are also highlighted in the plot. The low signal-to-noise, due to the drop in flux, combined with the emission lines, makes it impossible to detect individual absorption features.

Furthermore, in order to test the consistency of our model with the observed UV spectrum, we extrapolated our absorption corrected X-ray model to the UV region and calculated the predicted model flux at 1350 $\AA$ using our broken powerlaw parameters. We used the X-ray flux at 5 keV, F$_{5 keV}$~=~1.05~$\times$~10$^{-3}$~photons cm$^{-2}$ s$^{-1}$ keV$^{-1}$, and estimated the intrinsic continuum flux without absorption at 0.5 keV. Thus we calculated the UV flux to be F$_{1350~\AA}$~=~$\sim$~3.3~$\times$~10$^{-14}$ erg cm$^{-2}$ s$^{-1}$ $\AA^{-1}$, which overpredicts the observed flux by $\sim$~30~$\%$, but it is still consistent if we consider that we did not correct the X-ray intrinsic continuum flux at 5 keV to contributions from the reflected continuum.

\subsection{Comparison with Archival Datasets}

Based on the 2014 model we fitted the archival datasets of NGC 4151 from {\it Chandra}, {\it XMM-Newton}, and {\it Suzaku}. All of the final model parameters, including the 2014 and 2002 observations, can be found on Table 4. X-ray flux errors were calculated on XSPEC with a 68$\%$ confidence range. For all datasets the column density within X-High and D+Ea remain constant, except the 2000 {\it XMM-Newton}, which also exhibits the most different spectral shape (see Figure 1). It is important to note that the low flux states have column densities similar to 2014, while the high flux states show column densities like the 2002 model.
 
\bigskip

\begin{deluxetable}{lcc|cc}
\tablenum{3}
\tabletypesize{\small}
\tablecaption{Model Column Densities}
\tablewidth{0pt}
\tablehead{
\colhead{Ion} & \multicolumn{2}{c}{2014} & \multicolumn{2}{c}{2002} \\
\cline{2-5}
\colhead{} & \colhead{X-High} & \colhead{D+Ea} & \colhead{X-High} & \colhead{D+Ea}
}
\startdata
\ion{H}{1}   & 16.16 & 22.38 & 14.65 & 17.19 \\
\ion{He}{2}  & 17.48 & 21.08 & 15.98 & 20.61 \\
\ion{C}{1}   & -     & 14.34 & -     & -     \\
\ion{C}{2}   & -     & 19.10 & -     & -     \\
\ion{C}{4}   & -     & 17.98 & -     & 16.58 \\
\ion{N}{5}   & 15.75 & 17.04 & -     & 18.16 \\
\ion{O}{1}   & -     & 19.24 & -     & -     \\
\ion{O}{2}   & -     & 18.90 & -     & -     \\
\ion{O}{6}   & 17.21 & 17.81 & -     & 18.60 \\
\ion{O}{7}   & 19.20 & 17.49 & -     & 19.02 \\
\ion{Ne}{8}  & 17.65 & -     & 14.34 & 17.63 \\
\ion{Ne}{9}  & 18.55 & -     & 16.80 & 17.54 \\
\ion{Ne}{10} & 18.12 & -     & 17.91 & 16.21 \\
\ion{Mg}{9}  & 17.57 & -     & -     & 16.94 \\
\ion{Mg}{10} & 17.42 & -     & 14.16 & 16.28 \\
\ion{Mg}{11} & 17.80 & -     & 15.18 & 15.83 \\
\ion{Mg}{12} & 17.14 & -     & 16.13 & 14.14 \\
\ion{Si}{2}  & -     & 18.06 & -     & -     \\
\ion{Si}{3}  & -     & 16.37 & -     & -     \\
\ion{Si}{4}  & -     & 17.01 & -     & 15.57 \\
\ion{Si}{5}  & -     & 17.34 & -     & 17.10 \\
\ion{Si}{6}  & 14.18 & 16.73 & -     & 17.12 \\
\ion{Si}{7}  & 16.25 & 16.57 & -     & 17.45 \\
\ion{Si}{8}  & 17.37 & 15.53 & -     & 17.50 \\
\ion{Si}{9}  & 17.64 & -     & -     & 16.80 \\
\ion{Si}{10} & 17.66 & -     & 14.17 & 16.01 \\
\ion{Si}{11} & 17.32 & -     & 15.18 & 14.84 \\
\ion{Si}{12} & 17.00 & -     & 16.13 & 14.13 \\
\ion{Si}{13} & 17.09 & -     & 17.41 & -     \\
\ion{Si}{14} & 16.13 & -     & 17.65 & -     \\
\ion{Fe}{2}  & -     & 18.17 & -     & -     \\
\ion{Fe}{7}  & 14.88 & 16.77 & -     & 17.29 \\
\ion{Fe}{8}  & 16.75 & 16.65 & -     & 17.77 \\
\ion{Fe}{9}  & 17.63 & 15.22 & -     & 17.29 \\
\ion{Fe}{10} & 17.82 & -     & -     & 16.68 \\
\ion{Fe}{11} & 17.66 & -     & -     & 15.74 \\
\ion{Fe}{12} & 17.32 & -     & -     & 14.68 \\
\ion{Fe}{13} & 16.99 & -     & -     & -     \\
\ion{Fe}{14} & 16.59 & -     & -     & -     \\
\ion{Fe}{15} & 16.20 & -     & -     & -     \\
\ion{Fe}{16} & 16.02 & -     & 14.75 & -     \\
\ion{Fe}{17} & 16.66 & -     & 16.59 & -     \\
\ion{Fe}{18} & 16.31 & -     & 17.27 & -     \\
\ion{Fe}{19} & 15.69 & -     & 17.60 & -     \\
\ion{Fe}{20} & 14.54 & -     & 17.26 & -     \\
\enddata
\end{deluxetable}

The Fe K$\alpha$ equivalent width varies as a factor of $\sim$2 in high and low flux states, and the 2006 {\it{Suzaku}} dataset shows structure below 6.4 keV, which we initially attempted to fit with a broad Fe K$\alpha$ wing. However, it is more consistent with an emission feature around 6.08 keV that has been previously mentioned by \citet{2002ApJ...574L.123T}. It is statistically significant, but its source remains unclear. Addition of a broad wing did not show statistical improvement in other datasets. The unattenuated continuum contributions to the total flux varies from $\sim$~3$-$5~$\%$ in the high states to $\sim$~12$-$20~$\%$ in the low states.

The results from all datasets corroborate with the same overall model parameters, pointing to very stable intrinsic absorption. There are notable changes in the gas ionization states, which are approximately proportional to the flux variations. We also observe an interesting 'pattern' in the column density variations in the low flux epochs in contrast with the high flux ones, which we discuss in the next section. 

\floattable
\begin{deluxetable}{lcccccccccc} 
\tabletypesize{\small}
\tablewidth{0pt} 
\tablenum{4}
\tablecaption{Best Fitting Model Parameters}
\tablehead{
\colhead{} & \colhead{Observation} & \multicolumn{2}{c}{X-High} & \multicolumn{2}{c}{D+Ea} & \colhead{Reflection} & \colhead{Fe K$\alpha$ EW} & \colhead{D$_{f}$} & \colhead{Fit} \\
\cline{3-6}
\colhead{} & \colhead{} & \colhead{$\log$~N$_{H}$} & \colhead{$\log$~U} & \colhead{$\log$~N$_{H}$} & \colhead{$\log$~U} & \colhead{R} & \colhead{(eV)} & \colhead{($\%$)} & \colhead{$\chi$~$^{2}$}
}
\startdata 
\it{Chandra} & 2014 & $22.71^{+0.01}_{-0.01}$ & $0.42^{+0.04}_{-0.05}$ & 22.71 & $-0.81^{+0.14}_{-0.14}$ & $-2.17^{+0.14}_{-0.14}$ & 173$\pm$19 & 10$\%$ & 1568/1589 \\
             & 2000 & $22.77^{+0.02}_{-0.02}$ & $0.38^{+0.38}_{-0.18}$ & 22.77 & $-0.75^{+0.22}_{-0.21}$ & $-2.06^{+0.35}_{-0.35}$ & 209$\pm$13 & 12$\%$ & 415/813 \\
             & 2002 & $22.53^{+0.01}_{-0.01}$ & $1.09^{+0.05}_{-0.03}$ & 22.53 & $-0.31^{+0.03}_{-0.03}$ & $-0.82^{+0.09}_{-0.09}$ & 81$\pm$7 & 3$\%$ & 2578/2596 \\ 
             & 2002 & $22.48^{+0.01}_{-0.01}$ & $0.91^{+0.06}_{-0.06}$ & 22.48 & $-0.43^{+0.10}_{-0.10}$ & $-0.93^{+0.24}_{-0.24}$ & 132$\pm$6 & 5$\%$ & 417/421 \\
             & 2007 & $22.46^{+0.02}_{-0.02}$ & $0.60^{+0.05}_{-0.05}$ & 22.46 & $-0.68^{+0.08}_{-0.08}$ & $-0.49^{+0.17}_{-0.18}$ & 130$\pm$4 & 6$\%$ & 1828/1829 \\
\hline
\it{XMM-Newton} & 2000 & $22.76^{+0.01}_{-0.01}$ & $0.28^{+0.03}_{-0.03}$ & $22.48^{+0.03}_{-0.03}$ & $-0.98^{+0.34}_{-0.36}$ & $-2.57^{+0.09}_{-0.09}$ & 224$\pm$27 & 21$\%$ & 274/220 \\
\hline
\it{Suzaku} & 2006 & $22.76^{+0.01}_{-0.01}$ & $0.23^{+0.05}_{-0.06}$ & 22.76 & $-1.28^{+0.27}_{-0.26}$ & $-3.27^{+0.10}_{-0.10}$ & 282$\pm$38 & 17$\%$ & 1041/1017 \\        
\enddata
\tablecomments{{\it D$_{f}$ = Fraction of unattenuated continuum};  Reflection fraction R is given as negative values to indicate direct reflected continuum only. }
\end{deluxetable}

\section{Discussion}

We have created a well constrained model to account for the long term X-ray variability in NGC 4151. Based on the KRA2005 model, the absorption components respond to the incident radiation in terms of variations in ionization parameter, however we also observe an increase of 0.2 dex in the column density of the absorbers in the low flux epochs. The reflected continuum remains roughly constant over the 14 years that these observations span, becoming more dominant in lower flux states. The reflection fraction R varies from -3.29 to -0.49 in low and high state epochs. Such variability is consistent with a constant reflected flux and evidences a time lag, as the reflection fraction increases in response to a decrease in the intrinsic continuum flux. Based on our preliminary analysis, there is some variation on the Fe K$\alpha$ fluxes, however the sampling of the observations prevent a straightforward deconvolution of its response to continuum changes (we plan to explore this in more detail in the future). 

The unattenuated scattering continuum is also relatively constant, and its fraction with respect to the total flux is consistent with the scattering contributions to the UV spectrum, as seen in \citet{2001ApJ...551..671K}. The scattering photon index is greater than the intrinsic continuum photon index, $\Gamma$~=~2.4 compared to 2.3 below 0.5 keV, which could indicate that there are emission lines not taken into account, or possibly contributions from another source of continuum, i.e. a thermal bremsstrahlung component arising from the emission line gas \citep{1995MNRAS.275.1003W}. The presence of an uncovered fraction of the continuum in NGC 4151 was suggested by \citet{1998ApJS..114...73G}, in a study of the absorption of Seyfert 1 galaxies with ASCA. The X-ray flux of the ASCA observations was equivalent to our definition in this paper of high flux states, and the fraction of unocculted continuum corresponds to 3$-$4~$\%$ of the total flux. The 2000 {\it XMM-Newton} dataset is the only one with a different spectral shape, and it shows the most contribution from the uncovered continuum to the total flux. It is probably the only dataset in which transverse motion is seen, but variations in the ionization state dominate the spectral changes.

Given that there have been changes to the atomic data used by Cloudy since 2005, we generated the stability curve (S-curve) for the SED used in our models, illustrated in Figure 9. The S-curve shows us the regions of the gas stability to thermal perturbations; the lower ionization/temperature is dominated by line cooling, while highest ionization/temperature regions are dominated by Compton processes. The positions for X-High and D+Ea models are highlighted in the figure for both 2014 and 2002 {\it Chandra} observations. All of the absorption components lie in a quasi-stable region of the curve, with D+Ea closer to the line cooled, stable region, and X-High more susceptible to thermal perturbations by small changes in gas pressure. The gas pressure was computed by assuming that the absorption components were co-located, with a gas density of n$_{H(X-High)}$~=~10$^{6}$ cm$^{-3}$ and n$_{H(D+Ea)}$~=~1.6~$\times$~10$^{7}$ cm$^{-3}$. We obtained P$_{X-High}$~$\sim$~5.3$\times$10$^{-5}$ dyne cm$^{-2}$ and P$_{D+Ea}$~$\sim$~1.3$\times$10$^{-4}$ dyne cm$^{-2}$ for the 2014 model, while for 2002 P$_{X-High}$~$\sim$~2.8$\times$10$^{-4}$ dyne cm$^{-2}$ and P$_{D+Ea}$~$\sim$~3.6$\times$10$^{-4}$ dyne cm$^{-2}$. If they were in fact co-located, it would be possible for the absorbers to be in pressure equilibrium in the 2002 model, but not in 2014. 

Furthermore, we investigated the mechanisms of gas acceleration, and Figure 10 shows the predicted Force Multiplier (FM) as a function of ionization parameter for the 2014 and 2002 models, where FM is defined as the ratio of the total absorption cross-section, including bound-free and bound-bound transitions, to the Thomson cross-section. It is important to note that the FM values for D+Ea are slightly offset from the predicted curve, since they were generated using a filtered continuum. For a SMBH mass of M$_{BH}$~=~4.57~$\times~$10$^{7}~$~M$_{\odot}$ \citep{2006ApJ...651..775B}, we determined the Eddington luminosity, L$_{Edd}$~=~5.78~$\times$~10$^{45}$ erg s$^{-1}$. Based on our determination of the 2-10 keV source luminosity (L$_{2-10keV}$), and by assuming a bolometric correction of $\sim$~30 \citep{2001PASJ...53..647A}, we calculated the bolometric luminosity for our studied epochs. In order to radiatively drive the absorber, FM $\geq$ (L$_{Bol}$/L$_{Edd}$)$^{-1}$. Our results indicate that the source is radiating at an average of $\sim$~1~$\%$ $L_{Edd}$ in the low flux states, and around $\sim$1.5$-$2~$\%$ $L_{Edd}$ in the high flux states. At those rates, it would be necessary that FM~$\geq$~50 to cause radiatively driven outflows in high states, and FM~$\geq$~100 in low states. Thus, D+Ea could be radiatively driven in both high and low states, but X-High must be accelerated through a different process. 

Another possible mechanism for the outflows are the magnetohydrodynamic (MHD) disk winds, as more recently discussed by \citet{2010ApJ...715..636F} (and references therein), in which the absorbers are accelerated outwards along the magnetic field lines arising from the accretion disk.  In such models, the intrinsic absorption is seen as a continuous distribution of hydrogen column density per decade of ionization parameter, implying in a radially dependent ionized density of the form $n(r)$~$\propto$~r$^{-\alpha}$\citep{2014ApJ...780..120F}. An MHD flow would be characterized by $\alpha$~=~1, while a radiation driven wind would be represented by $\alpha$~=~2. Moreover, the ionization state of the ultra-fast outflows (UFO) suggested by \citet{2010A&A...521A..57T} are also consistent with a magnetically driven outflow.

\citep{2010A&A...521A..57T, 2011ApJ...742...44T, 2013MNRAS.430.1102T} discuss the mass outflows in terms of an unification of X-ray winds from UFOs to warm absorbers. A sample of Seyfert galaxies are classified as UFOs or Non-UFOs, with the UFOs being identified by highly ionized absorption features of \ion{Fe}{25}/\ion{Fe}{26} in the 7$-$8~keV band. Due to low effective area of {\it Chandra} detectors in that energy range, we could not detect an UFO in any of our observations. Interestingly, the Non-UFO absorbers are consistent with our model parameters for X-High, and may indicate a possible correlation between them. In NGC 4151, the UFO was observed in the 2006 {\it XMM-Newton} ObsID 0402660201 (illustrated in Figure 1), parameterized with with N$_{H}$~=~8~$\times$~10$^{22}$ cm$^{-2}$, $\log$~U$_{UFO}$~=~2.4, based on \citet{2013MNRAS.430.1102T}. We estimated that if located at r$_{UFO}$~=~1.3~$\times$~10$^{15}$ cm, the UFO would have a density of n$_{H}$~=~2.5~$\times$~10$^{8}$ cm$^{-3}$ (for more detailed work on the UFO parameters, see Kraemer et. al, in preparation).

\begin{figure}
\figurenum{9}
\epsscale{1.2}
\plotone{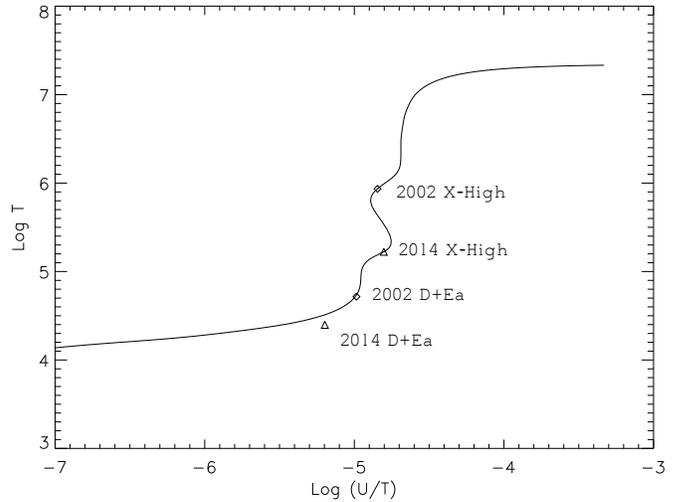}
\caption{Stability curve generated with photoionization models with the same SED for the X-ray absorbers. Vertical regions in the plot are quasi-stable, with negative slopes being unstable to thermal perturbations. The absorption components are represented as diamonds for the 2002 model, and triangles for 2014.
\label{fig:fig9}}
\end{figure}

\begin{figure}
\figurenum{10}
\epsscale{1.2}
\plotone{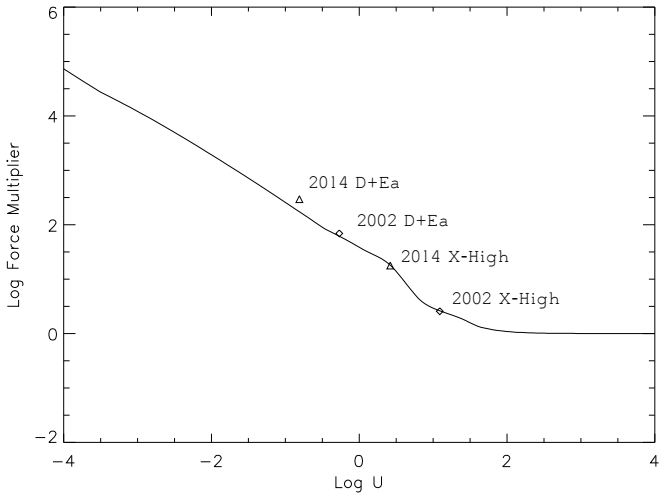}
\caption{Predicted Force Multiplier as a function of ionization parameter generated with photoionization models with the same SED for the X-ray absorbers. Diamonds correspond to the 2002 absorption components, and triangles represent the 2014 absorbers.
\label{fig:fig10}}
\end{figure}

We do not present a model fit for the 2006 {\it XMM} dataset in this paper, as we could not constrain the statistics of the fitting due to low number of filtered counts. However, we can estimate its model parameters in order to probe the magnetic nature of the outflows. Based on our results in Table 4 and a rough qualitative spectral fit, the 2006 observation is parameterized as N$_{H}$~$\sim$~5~$\times$~10$^{22}$ cm$^{-2}$, $\log$~U$_{X-High}$~$\sim$~0.5 and $\log$~U$_{D+Ea}$~$\sim$~$-$~0.7, which is consistent with a 2$-$10 keV flux of $\sim$~8~$\times$~10$^{-11}$ erg cm$^{-2}$ s$^{-1}$. Thus, we found that the density-radial dependence between the UFO and D+Ea was $n(r)$~$\propto$~r$^{-0.7}$, assuming that D+Ea is located at r$_{D+Ea}$~=~3.2~$\times$~10$^{17}$ cm, with n$_{H}$~=~6.3~$\times$~10$^{6}$ cm$^{-3}$ (see KRA2006). In a more general way, given the range in U$_{D+Ea}$ between high and low flux states, $\alpha$ can vary from 0.7~$\leq$~$\alpha$~$\leq$~0.9. As \citep{2014ApJ...780..120F} (and references therein) point out, MHD models are consistent with $\alpha$~$\sim$~1 up to 1.25, and therefore our results suggest that D+Ea is not part of the same MHD flow as the UFO. On the other hand, there is a very likely possibility that X-High is part of an MHD outflow. If that was the case, by considering U~$\propto$~1/($n(r)$r$^{2}$) and assuming $\alpha$~=~1, we can calculate the location of X-High based on the ionization parameter and location of the UFO. For the respective range in U$_{X-High}$ in high and low flux epochs, we found 2.5~$\times$~10$^{16}$~$\textless$~r$_{X-High}$~$\textless$~1~$\times$~10$^{17}$ cm. Moreover, such location for X-High is also consistent with the location of the Non-UFO absorbers (see \citealt{2012MNRAS.422L...1T,2013MNRAS.430.1102T}), which provides additional support to the fact that they may correspond to the same absorption component. This could indicate that at a sufficiently large radial distance, i.e. between X-High and D+Ea, there should be a break point between MHD-dominated and radiatively driven outflows.

Finally, the model results for the majority of the observations show a uniform column density within the absorbers. The intrinsic absorption respond to flux variations in terms of changes in its ionization state, but we also observe column density variations between high and low flux states. It is important to stress how well constrained are the column density variations, and how remarkably identical is the spectral shape of the 2000 and 2014 low flux {\it Chandra} datasets. Previous studies have discussed variabilities in the column density of D+Ea, such as the KRA2005 model, but changes in X-High makes us believe that there should be a fundamental and more universal explanation for the observed variations. We can think of many physically possible scenarios that could account for the column density variations, and the most straightforward answer is that simply there is more material in our line of sight. A decrease in the incident radiation in lower flux epochs could cause the gas not to be as efficiently driven as in higher flux ones. Another possible explanation is that variations in the ionizing radiation coming from the source induce changes in the gas condensation/evaporation ratios, affecting the gas stability and consequently making more material to condense from higher ionization states, as \citep{2001ApJ...561..684K} described in the multi-temperature wind models. Although it would be very difficult to quantitatively rule out these scenarios, they still rely on the coincidence that high and low flux epochs at some point return to nearly identical states.

With that in mind we suggest a third possibility, in terms of changes in the X-ray corona. In such scenario, we are looking through an unresolved column density distribution of gas, inversely proportional to its radial distance from the central AGN, such that the column density is higher closest to the source. The observed X-ray flux variabilities would then be due to a change in the average column density covering the source as the X-ray corona becomes less extended in lower flux states. This implies that the X-ray corona expands as the luminosity increases, and contracts closer to the black hole as the flux decreases again. What we obtain is an average column density, since we cannot deconvolve the column density distribution of the absorbers. Figure 11 shows an schematic representation of the model's geometry for NGC 4151. Furthermore, changes in the X-ray corona have been previously suggested by \citet{2014MNRAS.443.2746W} in the study of the flux variability of the Narrow Line Seyfert 1 galaxy 1H 0707$-$495. Given that the UV continuum and emission lines come from a more extended region than the X-ray corona, we would likely not observe this effect in the UV. 

\begin{figure*}
\centering
\includegraphics[scale=.4]{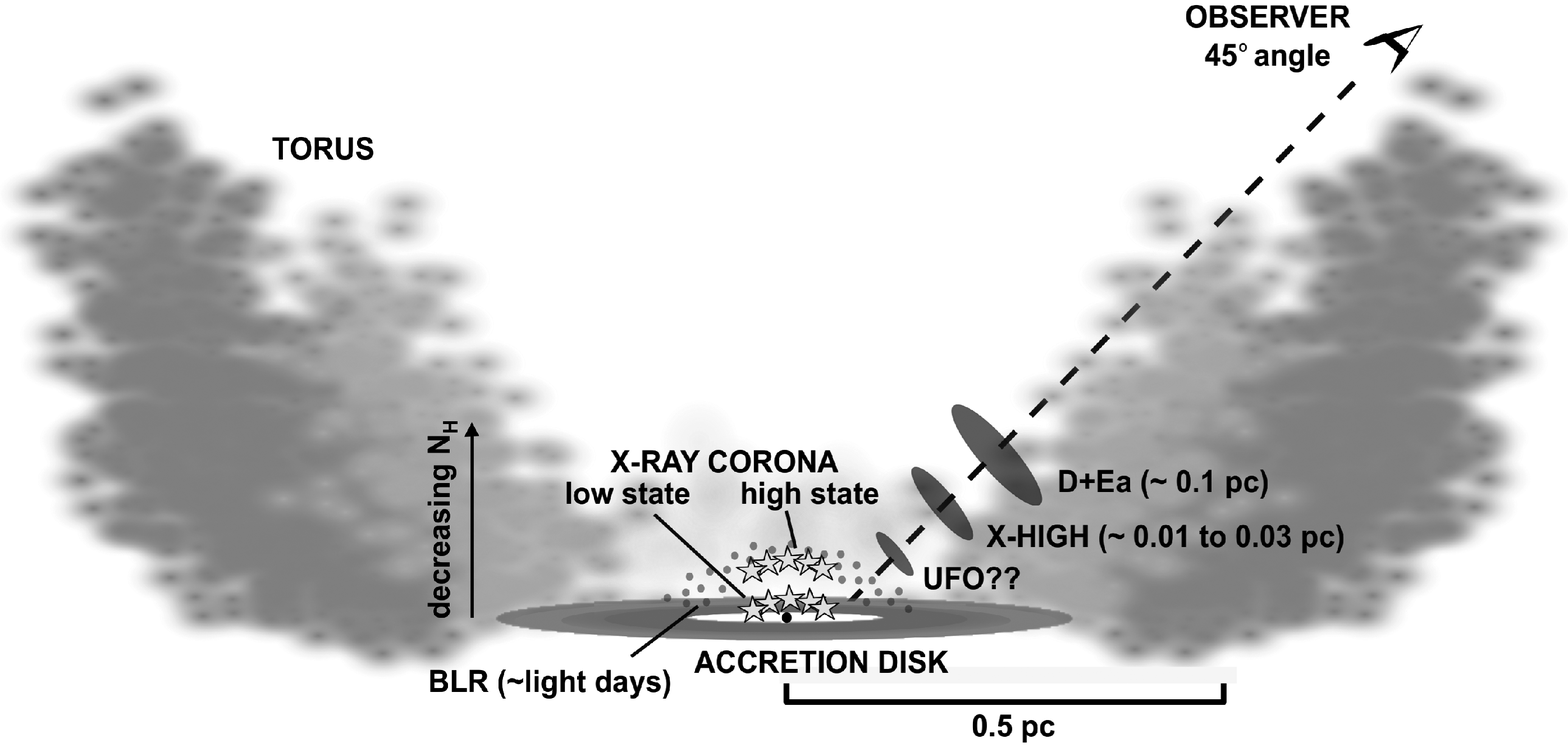}
\caption{Schematic representation, not to scale, of the accretion disk/AGN structure for NGC4151. Note the location of the X-ray absorbers along our line-of-sight. As the luminosity increases, the corona expands, and the average column density decreases because the true column density decreases with increasing polar angle
\label{fig:fig11}}
\end{figure*}

\section{Summary and Conclusions}
In this paper we revisit the analysis of the complex hard X-ray spectrum of NGC 4151 in order to address some still open, and yet fundamental questions about AGNs. We compared our model results for the 2014 simultaneous X-ray and UV observations of NGC 4151 with archival datasets from {\it Chandra}, {\it XMM-Newton}, and {\it Suzaku}. Over the past 14 years, we observed significant variability in the 2$-$10 keV flux of \objectname[NGC]{NGC 4151}. The intrinsic continuum is more heavily absorbed in the lower flux state than in the higher flux state epochs, but overall they present a very similar spectral shape. Our most notable results and concluding remarks are listed below:

1. The observations can be clearly divided into high and low states. Each group shows very distinct spectral characteristics, and the low states exhibit strong extended, unabsorbed emission at energies below 2 keV. The 2014 {\it Chandra} and {\it STIS} observations were found in a very low flux state. 

2. Our model results imply that the absorbers are very stable. Changes in the continuum are mostly intrinsic, meaning that the source is varying in brightness, rather than the flux is dropping due to the passage of optically thick gas into our line-of-sight. This is supported by the fact that the UV spectrum also showed significant flux variations, which ruled out the possibility of occultations. Furthermore, the remarkable similarities in the spectral shape of the 2014 and 2000 {\it Chandra} also suggest intrinsic variation. 

3. Additionally to the intrinsic continuum changes, the low flux state epochs show evidence of larger column densities in one or both of the absorbers. Among the possible explanations explored in detail in the previous section, we suggest that such variations in column density can be attributed to changes in X-ray coronal size. Specifically, the X-ray corona expands as the luminosity increases, and contracts as the flux decreases.

4. As suggested by KRA2005, MHD winds seem to play an important role in the mass outflows in NGC 4151, as X-High is consistent with being magnetically driven, along with the UFO absorber identified by \citet{2010A&A...521A..57T} on the 2006 {\it XMM-Newton} observation. Concurrently, D+Ea model parameters indicate that it cannot be part of the MHD flow, and it is likely accelerated through radiative processes. These results suggest that at a sufficiently large radial distance there should be a break point between MHD-dominated and radiatively driven outflows, which might help us understand the origin of the outflows and ultimately their role in AGN Feedback. 

5. The stability of the absorbers and the possibility of detection of a coronal expansion/contraction may be understood in the context of the high inclination (large polar angle) at which we are viewing NGC 4151. These effects may be not be as easily observable at lower inclination angles.

\acknowledgments
Support for this work was provided by the National Aeronautics and Space Administration through {\it Chandra} Award Number G04$-$15106X issued by the {\it Chandra X-ray Observatory Center}, which is operated by the Smithsonian Astrophysical Observatory for and on behalf of the National Aeronautics Space Administration under contract NAS8$-$03060, and from program GO$-$13508, support for which was provided by NASA through a grant from the Space Telescope Science Institute, which is operated by the Association of Universities for Research in Astronomy, Inc., under NASA contract NAS5-26555. This research has made use of data and/or software provided by the High Energy Astrophysics Science Archive Research Center (HEASARC), which is a service of the Astrophysics Science Division at NASA/GSFC and the High Energy Astrophysics Division of the Smithsonian Astrophysical Observatory. We thank Keith Arnaud, for his continuing maintenance and development of XSPEC, and Gary Ferland and associates, for the maintenance and development of Cloudy.

\end{document}